\documentclass[%
  aps,
  pra,
  reprint,
  superscriptaddress
]{revtex4-2}

\usepackage{graphicx}
\usepackage{bm}
\usepackage{physics}
\usepackage{mathtools}
\usepackage{amsfonts}
\usepackage{amssymb}
\usepackage{dcolumn}
\usepackage{comment}
\usepackage{booktabs}
\usepackage[mathlines]{lineno}
\usepackage{hyperref}
\usepackage{float}
\hypersetup{
colorlinks=true,
linkcolor=blue,
citecolor=blue,
urlcolor=blue
}

\begin{document}
\title{Source-Driven Magnetosonic Structures in Plasmas: Two-Fluid Modeling and PIC Validation}
\author{Prince Kumar},
\email{kumarprincephysics@gmail.com}
\author{S. K. Mishra} 
\affiliation{$^{1}$Planetary Sciences Division, Physical Research Laboratory, Ahmedabad, Gujarat 380009, India }
\date{\today}
\begin{abstract}
We develop and validate a one-dimensional electron–ion two-fluid model, self-consistently coupled to Maxwell’s equations, for magnetosonic structures generated by a moving charged source.
The model is independently benchmarked against particle-in-cell (PIC) simulations to assess its ability to reproduce electrostatic and electromagnetic responses. In the zero-magnetic-field limit, the electrostatic precursor structures are validated using the Korteweg--de Vries amplitude--width invariant. For a finite magnetic field, the frequency and propagation speed of the far wake are consistent with the low-frequency fast-magnetosonic branch, providing a linear benchmark for the electromagnetic response. The model reproduces the transition from a wake-dominated regime to localized forward precursors as the source velocity exceeds the magnetosonic speed, and to source-locked pinned structures at higher velocities. Independent PIC simulations recover the principal features of these regimes, supporting the validity of the two-fluid description for the parameters considered. 
By retaining separate electron and ion dynamics, electron inertia, and charge-separation effects, the validated model provides a framework for investigating source-driven plasma-wave structures across a broader range of plasma conditions and source–magnetic-field geometries, with potential applications to interpreting plasma-wave signatures associated with moving charged objects in space plasmas.
%
%
\end{abstract}
\maketitle
\section{Introduction}
\label{sec:introduction}
%

Moving objects can generate wakes and waves in a variety of physical environments, from ship-generated wakes in ordinary fluids to plasma disturbances associated with the Moon and orbiting spacecraft~\cite{reed2002ship,ogilvie1996observations,andre2021spacecraft}. In magnetized space environments, Jupiter's moon Io provides an example: its interaction with the Jovian plasma generates Alfvén wings and associated magnetic-field perturbations~\cite{kotsiaros2024juno}. Similarly, the interaction of the solar wind with Earth's magnetosphere produces a bow shock and a foreshock region characterized by plasma-wave activity~\cite{turc2023transmission}. These examples illustrate the broad relevance of waves and disturbances generated by interactions between bodies and their surrounding environments.
This broader context is particularly relevant to Earth's near-space environment, where the increasing population of artificial satellites, rocket fragments, and small orbital debris in low Earth orbit (LEO) has stimulated interest in alternative methods for detecting objects that are difficult to identify using conventional optical and radar techniques~\cite{leal2025managing,ocal2024inevitable,reynaud2026space}. A charged object moving through the ionospheric plasma perturbs the ambient electron and ion populations and can consequently excite collective plasma waves~\cite{sen2015,bernhardt2023observations,hughes2025observations}. The resulting plasma disturbance may extend over spatial scales much larger than the physical size of the object, providing a possible indirect signature for detecting and characterizing space objects. The relevance of this mechanism is not purely theoretical: in situ electric-field observations have reported plasma waves associated with charged space objects, including electromagnetic fluctuations whose amplitude, spectral content, and polarization are consistent with compressional Alfvén–magnetosonic disturbances~\cite{bernhardt2023observations}.

These nonlinear plasma disturbances can be electrostatic or electromagnetic, depending on plasma properties and magnetic field strength. A particularly intriguing class of these nonlinear responses comprises precursor structures that propagate ahead of the moving disturbance and pinned structures that remain localized and locked to it. The
general problem of waves generated by a moving source has a long history in fluid dynamics, where it
can generate both upstream- and downstream-propagating waves~\cite{maleewong2020transcritical,grimshaw2019transcritical,flamarion2022waves,strouhal1878besondere,lee1989experiments,rodrigues1987obstacle,kumar2026obstacle,prince}. In plasmas, this idea was
developed in the context of charged objects moving through an ionized
medium~\cite{sen2015}. Early theoretical studies showed that, in the absence of a magnetic field, a moving charged
source can generate electrostatic nonlinear perturbations in the form of
precursor and pinned solitons, in addition to the usual trailing wake. In the weakly
nonlinear and weakly dispersive limit, these structures can be described by
a forced Korteweg--de Vries (fKdV) equation, which provides a convenient
reduced description in terms of nonlinearity, dispersion, and external
forcing \cite{truitt2020, truitt2021, truitt2020simulating,sen2016,kumar2025study}.
The formation of these structures has also received experimental
support in flowing complex plasmas~\cite{jaswal2016, arora2021experimental}. However, recent studies have shown that electrostatic structures, particularly precursor structures, can be strongly damped under realistic ionospheric conditions, especially when the ion temperature is comparable to the electron temperature~\cite{resendiz2024ion,dharodi2026plasma, sam2025self,dharodi2026magnetic,sam2026ion,PrincePIC2026}. 
%
%
%
%

An electromagnetic mechanism provides a distinct route to their formation~\cite{scales2026investigation, sen2023electromagnetic, ganguli2025orbital}, and would not suffer such damping, making them longer-lived and easier to detect due to their larger spatial extent~\cite{bernhardt2023observations}.
A moving charged source can excite electromagnetic modes in a magnetized plasma, where magnetic forces provide the restoring mechanism for magnetosonic and Alfvénic disturbances.
 %
 Previous PIC simulations have demonstrated magnetosonic precursors and pinned solitons, while reduced inertial-MHD studies have investigated their nonlinear properties~\cite{kumar2020,acharya2022charged,acharya2021accelerated,sen2023electromagnetic}. Despite these advances, a time-dependent two-fluid model that evolves the electron and ion fluids separately, retains electron inertia and charge-separation effects, and couples their dynamics self-consistently to longitudinal and transverse electromagnetic fields, together with independent kinetic PIC validation, offers a complementary framework for investigating source-driven plasma responses across electrostatic and electromagnetic regimes.

In the present work, we, therefore, construct such a one-dimensional source-driven
electron--ion two-fluid model self-consistently coupled to Maxwell's
equations and independently validate its predictions against kinetic particle-in-cell (PIC) simulations. The moving object is represented by an externally prescribed
localized Gaussian charge source, allowing the source amplitude, spatial
width, and velocity to be controlled independently. 
We first use the
\(B_0=0\) limit as an electrostatic benchmark and verify the precursor
response. We then introduce a
finite magnetic field and identify the linear branch of the far wake through
its frequency spectrum and directly measured propagation velocity. The
nonlinear response is subsequently examined on both sides of the
magnetosonic threshold, thereby distinguishing a wake-dominated
subcritical regime from the forward localized precursor regime. A higher
source velocity is then used to identify a source-locked electromagnetic
pinned state through its co-moving density and transverse-field structure.
%

The present formulation provides a basis for extending the investigation to source–magnetic-field geometries beyond the one-dimensional configuration considered here and to a broader range of source characteristics. Such extensions could enable the study of oblique propagation and additional electromagnetic wave branches, including Alfvén and whistler waves, provided that the geometry, field components, and numerical resolution are adapted to their relevant dynamics. The validated two-fluid–Maxwell model could also be used to derive reduced descriptions, such as a forced Korteweg–de Vries equation, under appropriate weakly nonlinear and weakly dispersive conditions.

%
The remainder of this paper is organized as follows. Section~\ref{sec:fluidmodel} describes the one-dimensional electron--ion two-fluid model coupled self-consistently to Maxwell's equations and the numerical method used for its solution. Section~\ref{sec:fluidresults} presents the resulting electrostatic and electromagnetic responses, including the magnetosonic wake, precursor, and pinned structures. Section~\ref{sec:picmodel} introduces the kinetic particle-in-cell model and presents its validation against the two-fluid results for the magnetosonic wake, precursor, and pinned regimes. The main conclusions are summarized in Sec.~\ref{sec:conclusion}. 
%
\section{Two-Fluid Model for driven nonlinear magnetosonic strictures}
\label{sec:fluidmodel}
%
We consider a one-dimensional ($\partial/\partial y=\partial/\partial z=0$),
collisionless electron--ion plasma described by the two-fluid equations
coupled self-consistently with Maxwell's equations. Although all spatial
variations are restricted to the $x$ direction, both longitudinal and
transverse components of the fluid velocity and electromagnetic field are
retained. This allows the same model to describe electrostatic ion-acoustic
waves in the absence of a magnetic field and electromagnetic magnetosonic
waves when a background magnetic field is applied.
For each species $s=e,i$, the dimensional continuity and momentum equations
are
\begin{subequations}
\begin{align}
\frac{\partial n_e}{\partial t}
+\frac{\partial(n_e v_{ex})}{\partial x} &= 0, \\
\frac{\partial n_i}{\partial t}
+\frac{\partial(n_i v_{ix})}{\partial x} &= 0,
\label{eq:continuity}
\\
m_e\left(
\frac{\partial}{\partial t}
+v_{ex}\frac{\partial}{\partial x}
\right)v_{ex}
&=
-e(E_x+v_{ey}B_z)
-\frac{1}{n_e}\frac{\partial P_e}{\partial x},
\label{eq:electron_x}
\\
m_e\left(
\frac{\partial}{\partial t}
+v_{ex}\frac{\partial}{\partial x}
\right)v_{ey}
&=
-e(E_y-v_{ex}B_z),
\label{eq:electron_y}
\\
m_i\left(
\frac{\partial}{\partial t}
+v_{ix}\frac{\partial}{\partial x}
\right)v_{ix}
&=
e(E_x+v_{iy}B_z)
-\frac{1}{n_i}\frac{\partial P_i}{\partial x},
\label{eq:ion_x}
\\
m_i\left(
\frac{\partial}{\partial t}
+v_{ix}\frac{\partial}{\partial x}
\right)v_{iy}
&=
e(E_y-v_{ix}B_z).
\label{eq:ion_y}
\end{align}
\end{subequations}
The longitudinal electric field is obtained from Gauss's law (\ref{eq:gauss_fluid}),
while the transverse electromagnetic fields evolve according to

\begin{subequations}
\begin{align}
\frac{\partial E_x}{\partial x}
&=
\frac{e}{\epsilon_0}
\left(n_i-n_e+n_b\right),
\label{eq:gauss_fluid}
\\
\frac{\partial B_z}{\partial t}
&=
-\frac{\partial E_y}{\partial x},
\qquad
\frac{\partial E_y}{\partial t}
=
-c^2\frac{\partial B_z}{\partial x}
-\frac{J_y}{\varepsilon_0}
\label{eq:maxwell_fluid}
\end{align}
\end{subequations}
where the transverse current density is
$J_y=e(n_i v_{iy}-n_e v_{ey})$. The pressure is closed using a
polytropic equation of state,
$P_s=P_{s0}(n_s/n_0)^\gamma$, with $s=e,i$.
The moving charged object is represented by a localized Gaussian density
source,
%
$n_b(x,t)
=
n_{b0}
\exp\left[
-\frac{(x-V_b t)^2}{2\sigma_b^2}
\right],$
%
where $n_{b0}$, $\sigma_b$, and $V_b$ denote the source amplitude, spatial
width, and propagation velocity, respectively. This prescription follows the
effective-source approach commonly used in the precursor-soliton literature
and provides a controlled way of comparing the electrostatic and
electromagnetic responses.
For numerical calculations, the variables are normalized to the electron
plasma scales according to $x=x/d_e$, $t=\omega_{pe}t$,
$n_s=n_s/n_0$, $\bm{v}_s={v}_s/c$,
${E}=e{E}/(m_ec\omega_{pe})$, and
${B}=e{B}/(m_e\omega_{pe})$, 
%
where dimensional variables, $n_0$ is the equilibrium plasma density, $c$ is the speed of light, $e$ is the elementary charge, $m_e$ is the electron mass, and $\epsilon_0$ is the vacuum permittivity. The characteristic electron inertial length is defined as $d_e = c/\omega_{pe}$, where $\omega_{pe} = (n_0 e^2/\epsilon_0 m_e)^{1/2}$ is the electron plasma frequency.
In normalized form, the governing system can be written compactly as
\begin{subequations}
\begin{align}
\frac{\partial n_e}{\partial t}
+\frac{\partial (n_e v_{ex})}{\partial x}
&=0,
\label{eq:ne}
\\[1mm]
\frac{\partial n_i}{\partial t}
+\frac{\partial (n_i v_{ix})}{\partial x}
&=0,
\label{eq:ni}
\\[1mm]
\frac{\partial v_{ex}}{\partial t}
+v_{ex}\frac{\partial v_{ex}}{\partial x}
&=
-(E_x+v_{ey}B_z)
-\frac{\beta_e}{n_e}
\frac{\partial n_e^\gamma}{\partial x},
\label{eq:vex}
\\[1mm]
\frac{\partial v_{ey}}{\partial t}
+v_{ex}\frac{\partial v_{ey}}{\partial x}
&=
-(E_y-v_{ex}B_z),
\label{eq:vey}
\\[1mm]
\frac{\partial v_{ix}}{\partial t}
+v_{ix}\frac{\partial v_{ix}}{\partial x}
&=
\mu(E_x+v_{iy}B_z)
-\frac{\mu\beta_i}{n_i}
\frac{\partial n_i^\gamma}{\partial x},
\label{eq:vix}
\\[1mm]
\frac{\partial v_{iy}}{\partial t}
+v_{ix}\frac{\partial v_{iy}}{\partial x}
&=
\mu(E_y-v_{ix}B_z),
\label{eq:viy}
\\[1mm]
\frac{\partial E_x}{\partial x}
&=
n_i-n_e+n_b,
\label{eq:Ex}
\\[1mm]
\frac{\partial B_z}{\partial t}
&=
-\frac{\partial E_y}{\partial x},
\label{eq:Bz}
\\[1mm]
\frac{\partial E_y}{\partial t}
&=
-\frac{\partial B_z}{\partial x}
-\left(n_i v_{iy}-n_e v_{ey}\right).
\label{eq:Ey}
\end{align}
\label{eq:normalized_fluid}
\end{subequations}
where $\mu=m_e/m_i$, $\beta_e=v_{te}^2/c^2$, and
$\beta_i=v_{ti}^2/c^2$.

These coupled two-fluid equations are solved numerically using a Fourier pseudo-spectral method in the spatial coordinate~\cite{coutsias1989spectral}. The spatial domain is discretized using \(N=2048\) uniformly spaced grid points, and the spatial derivatives are evaluated in Fourier space. 
The resulting system of first-order ordinary differential equations is advanced in time using a fourth-order Runge--Kutta (RK4) scheme with a time step \(\Delta t=0.005\).  For simplicity and to reduce simulation time, we consider the ion-to-electron mass ratio ($\mu$) to be 25 in the paper. 
%
The moving charged source propagates along the \(x\)-direction and is represented by a Gaussian density perturbation. In all cases presented in the paper, we consider the source amplitude $n_{b0} = 0.1$ and the width $\sigma_{b} = 3.0$.
As the source moves through the plasma, it perturbs the local balance between the electron and ion densities. The resulting charge imbalance establishes a longitudinal electric field \(E_x\) through Poisson's equation. This field accelerates the electrons and ions in opposite directions, while their finite inertia and the pressure-gradient terms determine their longitudinal response. The associated density variations evolve according to the continuity equations, continually modifying the charge distribution and the longitudinal electric field.
For a finite background magnetic field, the Lorentz force couples the longitudinal and transverse motions of each species. The resulting transverse velocities generally differ between electrons and ions, giving rise to a net transverse current,
$j_y=n_i v_{iy}-n_e v_{ey}.$
This current, together with the spatial variation of the magnetic perturbation \(B_z\), governs the evolution of the transverse electric field \(E_y\) through Amp\`ere's law. Faraday's law couples \(E_y\) to the evolution of \(B_z\), thereby establishing a self-consistent electromagnetic response to the source-driven disturbance.
The electromagnetic response also couples back to the longitudinal dynamics through the magnetic-force terms in the \(x\)-momentum equations, which are proportional to \(v_{ey}B_z\) for electrons and \(v_{iy}B_z\) for ions, with the signs determined by the respective species equations. These forces modify the longitudinal accelerations, thereby influencing the subsequent density evolution and charge separation. The source-driven disturbance thus evolves through the coupled dynamics of the longitudinal electric field, electron and ion motion, transverse current, and electromagnetic fields. 
\section{Results and Discussion}
\label{sec:fluidresults}
%
\subsection{Electrostatic precursor: zero-magnetic-field benchmark}
\label{subsec:electrostatic}
\begin{figure}[t]
    \centering
    \includegraphics[width=\columnwidth]{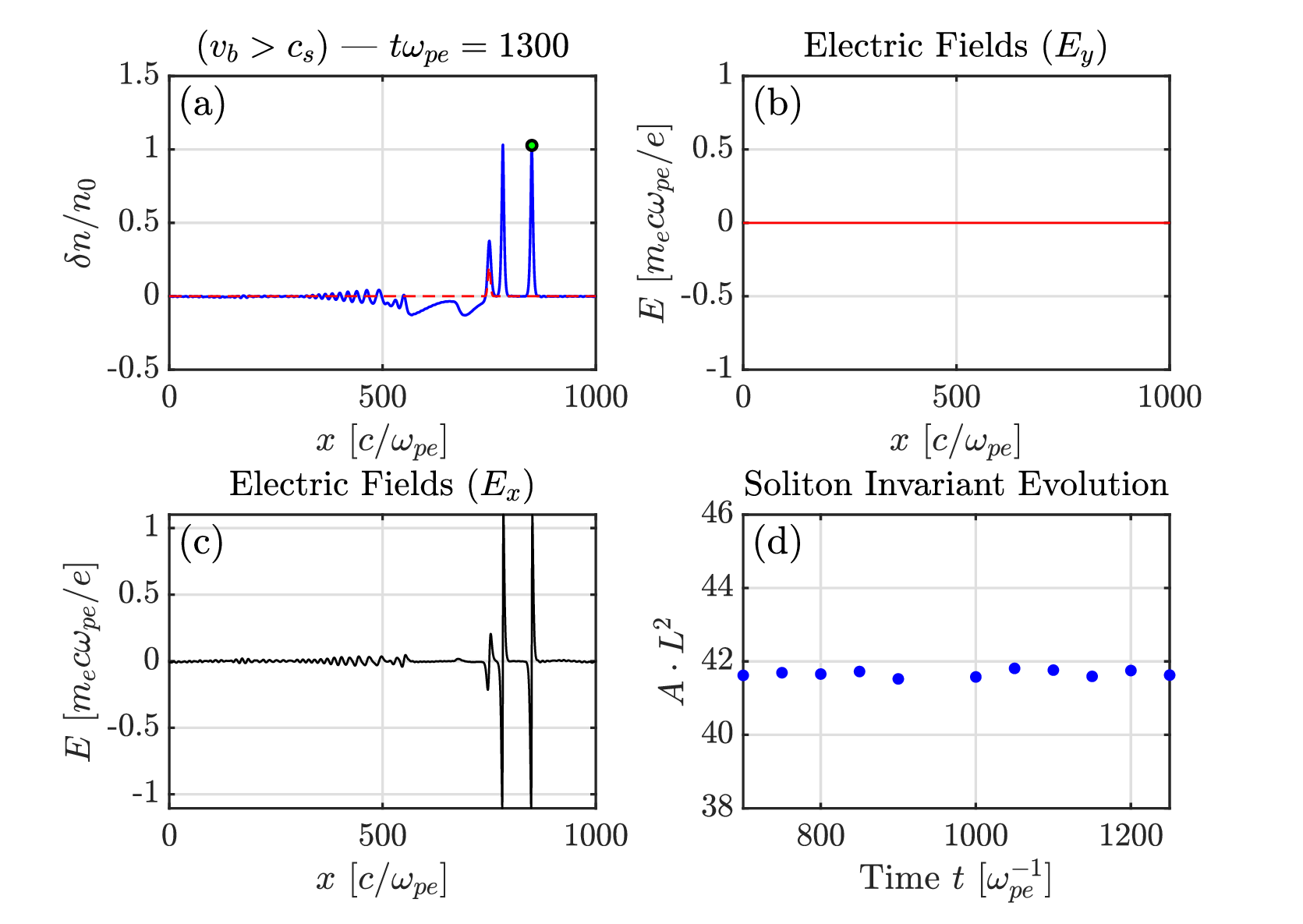}
    \caption{
    Electrostatic precursor benchmark for $B_0=0$ and a superacoustic source
    velocity. (a) Normalized density perturbation
    $\delta n/n_0$ showing the localized precursor pulses ahead of the moving
    source; (b) transverse electric field $E_y$; (c) longitudinal electric
    field $E_x$; and (d) temporal evolution of the soliton invariant
    $A L^2$, where $A$ is the normalized density amplitude and $L$ is the
    characteristic pulse width.
    }
    \label{fig:fluid_es}
\end{figure}

We first consider the limiting case $B=0$ in order to isolate the
electrostatic response and to benchmark the two-fluid model against the
established ion-acoustic precursor picture. In this limit, the transverse
perturbation is absent, and the nonlinear response is dominated by
the longitudinal electric field $E_x$. For a source moving faster than the unperturbed ion-acoustic speed,
\(V_b>V_{\rm cs}\), where \(V_{\rm cs}=\sqrt{k_B T_e/m_i}=v_{th,e}/\sqrt{m_i/m_e}\),
the source-driven compression cannot propagate away from the source as a
linear ion-acoustic disturbance, leading to nonlinear steepening at the
front of the source. 
It can be seen from  Fig.~\ref{fig:fluid_es} (a) that the moving source at a speed greater than the ion acoustic speed generates the precursor structures, followed by their validation, as presented in Fig.~\ref{fig:fluid_es} (d), which is shown by the constant nature of the leading soliton parameter $AL^2$ ($A$ and $L$ are the amplitude and width of the soliton, respectively).
As shown in Fig.~\ref{fig:fluid_es} (b), the transverse electric field is
zero throughout the domain, whereas the longitudinal component, as presented in Fig.~\ref{fig:fluid_es} (c),
exhibits pronounced localized structures in the region ahead of the source.
The absence of a significant $E_y$ component confirms that the $B_0=0$ response is electrostatic. 
Physically, the localized source enters the longitudinal field
equation through the Poisson equation. The resulting electric field
accelerates electrons and ions differently, producing charge
redistribution and density compression through their respective
continuity equations. With $B_0=0$ and the transverse fields initially
absent, the magnetic coupling that drives transverse species motion
is absent in this calculation, consistent with the negligible
$E_y$ in Fig.~\ref{fig:fluid_es}(b). 
%
%
\subsection{Electromagnetic precursor and pinned structures}
\label{subsec:em_precursor}
\subsubsection{Validation of electromagnetic solitons}
\label{subsec:em_validation}
\begin{figure}[t]
    \centering
    \includegraphics[width=\columnwidth]{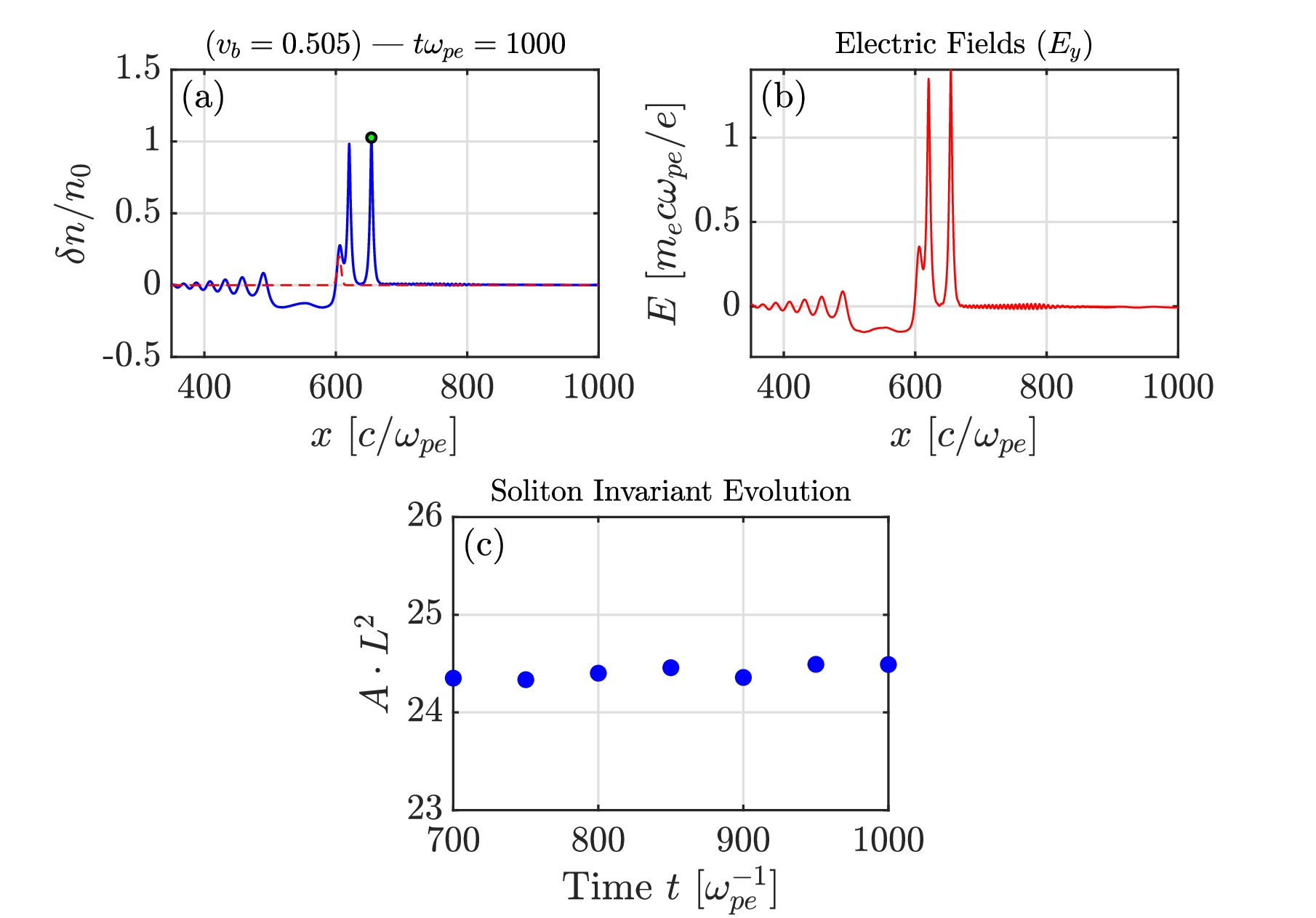}
    \caption{
    Validation of the electromagnetic precursor soliton in the two-fluid
    model for \(V_b=0.505c>V_{\rm ms}\). (a) Normalized density perturbation
    \(\delta n/n_0\), showing nonlinear localized precursor structures ahead
    of the moving source. (b) Transverse electric field \(E_y\) associated
    with the localized electromagnetic structures. (c) Temporal evolution
    of the soliton invariant \(A L^2\), which remains approximately constant,
    providing a KdV-type amplitude--width diagnostic.
    }
    \label{fig:emsol}
\end{figure}

%
%
%
Our next target is to validate the electromagnetic precursor generated
in the presence of a finite ambient magnetic field.
In contrast to the
electrostatic case, the source excites coupled longitudinal and transverse
field components, and the resulting localized structures
belong to the electromagnetic magnetosonic branch. Figure~\ref{fig:emsol} shows the corresponding electromagnetic precursor structure for \(V_b=0.505c\), which is above the magnetosonic speed \(V_{\rm ms}\).  For the cold two-fluid parameters
$m_i/m_e=25$ and $B_0=2.5$ used here, the 
magnetosonic speed is $V_{\rm ms}\simeq0.4402c$, as presented in Appendix~\ref{app:linear_dispersion}. Therefore, the Mach number for these parameter values can be written as  \(M=V_b/V_{\rm ms}\simeq 1.14\)
%
%
In Fig.~\ref{fig:emsol}(a), the normalized density perturbation
\(\delta n/n_0\) exhibits a sequence of oscillations in the far wake,
followed by a set of localized, finite-amplitude compressive structures
ahead of the moving source. 
%
The localization of these structures on the forward side of the source is
the characteristic precursor configuration obtained when the source moves
slightly faster than the relevant magnetosonic propagation speed.
The corresponding transverse electric field \(E_y\), shown in
Fig.~\ref{fig:emsol}(b), exhibits sharply localized pulses at essentially
the same spatial locations as the leading density structures. The
simultaneous appearance of \(\delta n\) and \(E_y\) is important: the
localized excitation is therefore not purely electrostatic, but involves a
coupled longitudinal--transverse electromagnetic response. This provides
the finite-\(B_0\) counterpart of the electrostatic benchmark.

A further validation is provided by the temporal evolution of the
amplitude--width invariant in Fig.~\ref{fig:emsol}(c). Over the interval
\(700\lesssim t\omega_{pe}\lesssim1000\), the quantity
$A L^2$
remains approximately constant, with only a small temporal variation around
\(A L^2\simeq24.4\). Such weak variation indicates that the leading
localized structure retains an approximately invariant amplitude--width
relation during its evolution. 
%
%
%
To identify the relevant linear wave branch associated with these nonlinear structures, we
next examine the subcritical and supercritical source-velocity regimes and
diagnose the propagation properties of the far wake.
\subsubsection{Magnetosonic character of the far wakes}
\label{subsec:wake}
\begin{figure}[t]
    \centering
    \includegraphics[width=\columnwidth]
    {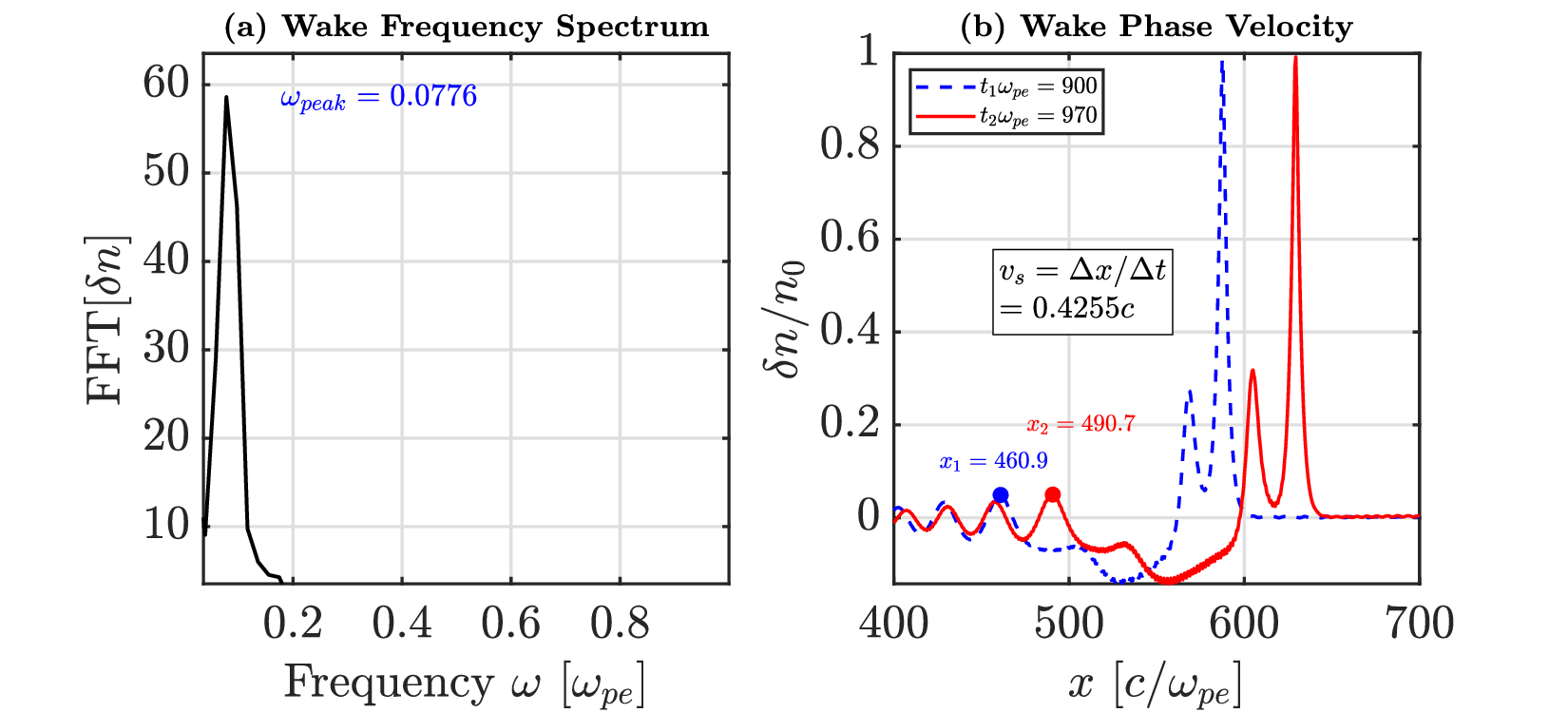}
    \caption{
    Diagnostics of the far-wake perturbation in the magnetized
    two-fluid model. (a) Temporal Fourier spectrum of the density
    perturbation, showing a dominant peak at
    $\omega_{\rm peak}\simeq0.0776\,\omega_{pe}$.
    (b) Density profiles at $t\omega_{pe}=900$ and $970$.
    Tracking a representative far-wake maximum gives
    $v_{\rm ph}\simeq0.4255c$. The inferred effective wavenumber
    and wavelength are $k_{\rm eff}\simeq0.1824\,\omega_{pe}/c$
    and $\lambda_{\rm eff}\simeq34.45\,c/\omega_{pe}$,
    respectively.
    }
    \label{fig:wake}
\end{figure}

To identify the linear wave branch underlying the nonlinear
structures discussed above, we examine the far-wake region, where
the perturbation amplitude is relatively small and a linear-wave
interpretation is more appropriate. Two complementary diagnostics
are presented in Fig.~\ref{fig:wake}: the temporal Fourier spectrum
of the density perturbation and the propagation of a representative
density maximum.
As shown in Fig.~\ref{fig:wake}(a), the wake spectrum exhibits a
dominant low-frequency peak at
$\omega_{\rm peak}\simeq0.0776\,\omega_{pe}$, consistent with a
slow magnetized-plasma response. The propagation speed is estimated
from the displacement of a density maximum between
$t_1\omega_{pe}=900$ and $t_2\omega_{pe}=970$
[Fig.~\ref{fig:wake}(b)]. The maximum shifts from
$x_1\simeq460.9\,c/\omega_{pe}$ to
$x_2\simeq490.7\,c/\omega_{pe}$, yielding
$v_{\rm ph}=\frac{x_2-x_1}{t_2-t_1}
\simeq0.4255c$ .
Combining the measured frequency and propagation speed yields
$k_{\rm eff}=\frac{\omega_{\rm peak}}{v_{\rm ph}}
\simeq0.1824\,\frac{\omega_{pe}}{c},$
$\lambda_{\rm eff}=\frac{2\pi}{k_{\rm eff}}
\simeq34.45\,\frac{c}{\omega_{pe}}.$
%
These measurements are consistent with the analytical linear magnetosonic dispersion relation given in Appendix~\ref{app:linear_dispersion}.
For the cold two-fluid parameters
$m_i/m_e=25$ and $B_0=2.5$ used in
Appendix~\ref{app:linear_dispersion}, the low-wavenumber
magnetosonic speed is $V_{\rm ms}\simeq0.4402c$, close to the
value measured here in the wake ($\simeq 0.4255c$). Moreover, the analytical dispersion relation
predicts $\omega\simeq0.0760\,\omega_{pe}$ at
$k_x\simeq0.1824\,\omega_{pe}/c$, approximately $2.1\%$ below
the measured spectral peak reported here.
Taken together, the measured frequency and
propagation speed, alongside their consistency with the cold
two-fluid dispersion relation, support identifying the far wake
as a low-frequency magnetosonic disturbance. 
%
%
%
The fact that the measured propagation
velocity approaches \(V_{\rm ms}\) supports the interpretation of the
far-wake oscillations as fast magnetosonic disturbances. This provides a
linear reference state for the nonlinear structures studied subsequently:
the precursor and pinned structures can then be viewed as nonlinear
magnetosonic responses generated from the same underlying electromagnetic
plasma dynamics.
%
\subsubsection{Magnetosonic Precursor solitons}
\label{subsec:mach}
\begin{figure}[t]
    \centering
    \includegraphics[width=\columnwidth]{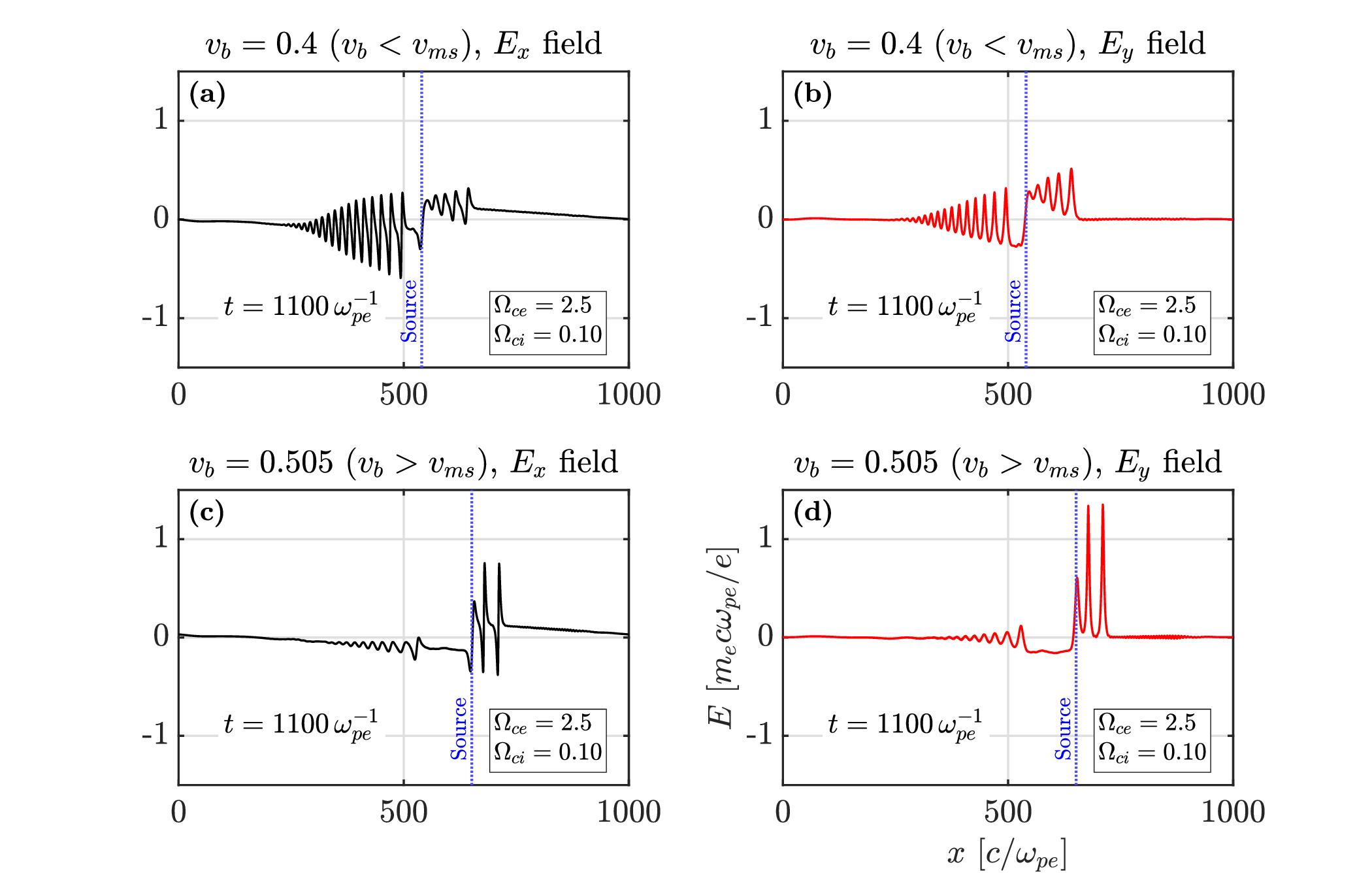}
    \caption{
    Longitudinal and transverse electric-field structures for two source
    velocities in the magnetized two-fluid model. The upper row corresponds
    to $V_b=0.4c<V_{\rm ms}$~($M = 0.90$), while the lower row corresponds to
    $V_b=0.505c>V_{\rm ms}$~($M \simeq 1.14$). Panels (a) and (c) show $E_x$, and panels
    (b) and (d) show $E_y$. The vertical dotted lines mark the source
    position at time $t\omega_{pe} = 1100$.
    }
    \label{fig:mach}
\end{figure}

In this section, we characterise the precursor solitons at different source velocities. 
In this case, both electrons and ions are assumed to be cold, with $\beta_i=\beta_e=0.0001$. We consider two cases: (i) a sub-magnetosonic source with $V_b=0.4c<V_{\rm ms}$, and (ii) a slightly super-magnetosonic source with $V_b=0.505c>V_{\rm ms}$. 
Figure~\ref{fig:mach} illustrates the transition in the field response as the source velocity crosses the magnetosonic propagation speed. The source moves in the direction of increasing \(x\), and the vertical dotted line indicates its instantaneous position. The two rows correspond to the subcritical case, \(V_b=0.4c<V_{\rm ms}\), and the weakly supercritical case, \(V_b=0.505c>V_{\rm ms}\), while the columns display the longitudinal electric field \(E_x\) and the transverse electric field \(E_y\), respectively. 

In the subcritical case, \(V_b=0.4c<V_{\rm ms }=0.4402c\), panel~(a) shows that the longitudinal electric field \(E_x\) is dominated by a spatially extended oscillatory response (wakes) concentrated predominantly behind the source, with a comparatively weak perturbation in front of the source. 
This indicates that the source motion is unable to outrun the characteristic magnetosonic disturbances of the background plasma. Consequently, the energy deposited by the moving source is carried primarily into the trailing region, producing a wake-like response rather than a well-defined forward-propagating precursor. 
%
The corresponding transverse electric field in panel~(b) exhibits a qualitatively similar subcritical response. The \(E_y\) perturbation consists of a finite-amplitude oscillatory packet extending predominantly on the trailing side of the source, while the field ahead of the source remains comparatively weak. The simultaneous presence of \(E_x\) and \(E_y\) demonstrates that the disturbance is electromagnetic rather than purely electrostatic. Nevertheless, the absence of a pronounced localized forward structure is consistent with the sub-magnetosonic character of the source motion. Thus, panels~(a) and~(b) together show that, for \(V_b<V_{\rm ms}\), the dominant response is a source-driven wake with the largest field activity occurring in the region traversed by the source.

A qualitatively different structure appears when the source velocity is increased slightly above the magnetosonic speed, \(V_b=0.505c>V_{\rm ms} = 0.4402\). In panel~(c), corresponding to \(E_x\), the wake-like oscillations are substantially weaker away from the source, whereas a pronounced localized field structure develops on the forward side of the source. Because the source moves toward increasing \(x\), the region to the right of the vertical dotted line corresponds to the fore-wake region. The emergence of a localized perturbation in this region is therefore direct evidence that the disturbance can propagate ahead of the moving source once \(V_b\) exceeds the magnetosonic phase speed. The field amplitude is considerably more concentrated spatially than in the subcritical case, indicating a transition from a predominantly radiative wake response toward a nonlinear, spatially localized precursor structure.
The transverse field in panel~(d) provides the corresponding electromagnetic signature of this transition. A strong, sharply localized \(E_y\) structure develops ahead of the source, with an amplitude substantially larger than the weak oscillatory perturbations present in the subcritical case.
In contrast to the subcritical case, where the dominant response remains behind the source, the supercritical case supports a coherent disturbance in front of the source because the source velocity exceeds the characteristic magnetosonic propagation speed.

%
%
\subsubsection{Pinned electromagnetic soliton}
The precursor results discussed above show that the nonlinear response
changes substantially as the source velocity approaches and exceeds the
characteristic magnetosonic speed. To examine the higher-velocity
regime, we next consider a source propagating at
$V_b=0.7c$, while keeping the background magnetic field,
source amplitude, and source width unchanged from the preceding
magnetosonic calculations. With $V_{\rm ms}\simeq0.4402c$, this corresponds to
a magnetosonic Mach number $M\simeq1.59$.
In contrast to the freely propagating precursor structures obtained
near the magnetosonic threshold, the nonlinear response at this higher
source velocity remains spatially localized around the moving source,
indicating the formation of a pinned magnetosonic structure.

\begin{figure}[t]
    \centering
    \includegraphics[width=\columnwidth]{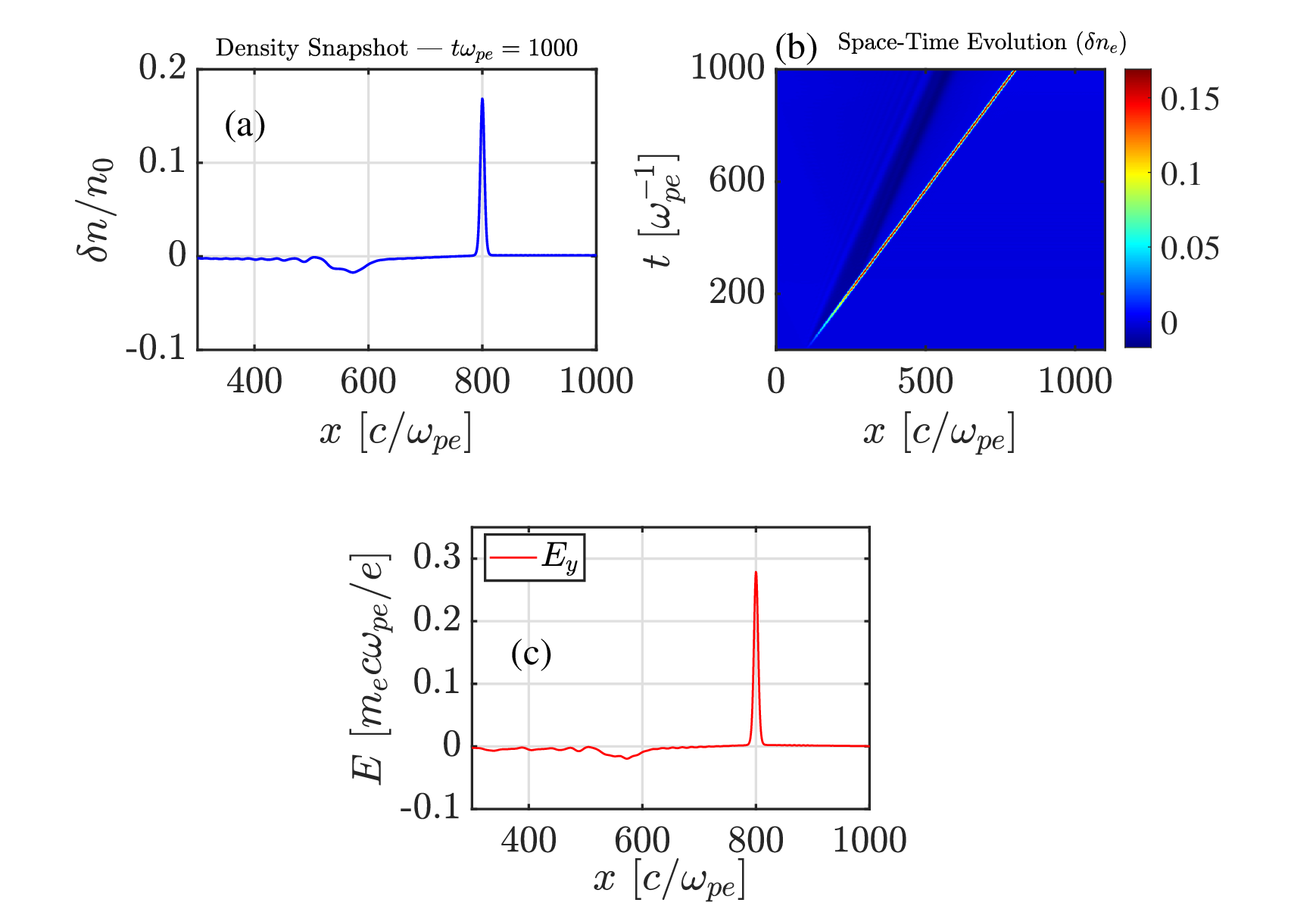}
    \caption{
    Electromagnetic pinned structure from the two-fluid model for
    $V_b=0.7c$ ($M=1.59$). (a) Density perturbation
    $\delta n/n_0$ at $t\omega_{pe}=1000$; (b) space--time evolution
    of the density perturbation, showing the co-moving structure along
    the source trajectory; and (c) transverse electric field $E_y$.
    }
    \label{fig:fluid_pinned}
\end{figure}

Figure~\ref{fig:fluid_pinned} summarizes the resulting two-fluid
solution. Panel (a) shows the density perturbation at
$t\omega_{pe}=1000$. A pronounced localized density enhancement is
observed near $x\simeq800\,c/\omega_{pe}$, with a peak amplitude of
approximately $\delta n/n_0\simeq0.16$. The position of this density
maximum coincides, to within the spatial resolution of the simulation,
with the instantaneous source position. Thus, unlike the precursor
pulses discussed previously, the dominant nonlinear density structure
does not detach from the source and propagate into the forward region.
A weaker density depression extends behind the localized peak,
reflecting the accompanying wake response, but the strongest nonlinear
compression remains centered on the source.

The propagation of the density perturbation is shown more directly in
the space--time diagram of panel (b). The colored line denotes the
trajectory of the moving source, $x_b(t)=x_{b0}+V_b t$. The localized
density structure remains spatially attached to this trajectory
throughout the simulation: its ridge follows the source with the same
slope in the $(x,t)$ plane rather than separating from it as an
independently propagating precursor. In a frame moving with the
source, the structure is therefore approximately stationary. This
co-moving behavior provides the defining signature of a pinned
soliton.
%
The transverse electric field is shown in panel (c). A narrow,
high-amplitude peak in $E_y$ is localized at approximately the same
position as the density maximum, reaching an amplitude of about
$E_y\simeq0.27$ in the present normalization. The strong localization
of the transverse field together with the density enhancement
demonstrates that the structure is not simply a stationary density
compression. Instead, it involves a coupled electromagnetic response
of the plasma. The much smaller oscillatory field fluctuations away
from the source further emphasize the localization of the nonlinear
structure around the source trajectory.
The transition from the precursor regime to the pinned state can thus
be viewed in terms of the relative motion between the nonlinear plasma
structure and the driving source. At $V_b\lesssim V_{\rm ms}$, the
response is dominated by wake oscillations with only weak forward
structures, while near the magnetosonic threshold pronounced
precursor pulses develop ahead of the source. At the larger velocity
considered here, $V_b=0.7c$, the dominant nonlinear disturbance no
longer separates from the source but remains attached to it. The
resulting structure therefore belongs to the pinned branch of the
electromagnetic nonlinear response.

%
%
\section{Kinetic PIC Model for driven magnetosonic soliton in plasmas}
\label{sec:picmodel}

To validate the fluid predictions at the kinetic level, we developed a
one-dimensional electromagnetic particle-in-cell (PIC) model~\cite{birdsall1991}. The PIC model
is one-dimensional in configuration space, while the particles retain the
two velocity components $(v_x,v_y)$, and the fields contain
$(E_x,E_y,B_z)$. The plasma consists of electrons and ions with
$q_e=-e$, $q_i=e$, and a reduced mass ratio $m_i/m_e=25$.
%
%
%
The kinetic evolution is represented by the one-dimensional Vlasov equation
for each species $s$,
\begin{equation}
\begin{aligned}
\frac{\partial f_s}{\partial t}
+v_x\frac{\partial f_s}{\partial x}
+\frac{q_s}{m_s}
\Bigg[
&(E_x+v_yB_z)\frac{\partial f_s}{\partial v_x}
\\
&+(E_y-v_xB_z)\frac{\partial f_s}{\partial v_y}
\Bigg]
=0 ,
\end{aligned}
\label{eq:vlasov_pic}
\end{equation}
where $f_s(x,v_x,v_y,t)$ is the distribution function. In the PIC
representation, the distribution functions are sampled by computational
macroparticles whose trajectories satisfy the Lorentz force,
$d\bm{p}_s/dt=q_s(\bm{E}+\bm{v}_s\times\bm{B})$, with
$\bm{E}=(E_x,E_y,0)$ and
$\bm{B}=(0,0,B_0)$. The particle trajectories are
advanced with the Boris algorithm~\cite{boris1970relativistic}.

For the one-dimensional electromagnetic geometry, the field evolution is
given in normalized units by
\begin{equation}
\frac{\partial\delta B_z}{\partial t}=-\frac{\partial E_y}{\partial x},\;
\frac{\partial E_y}{\partial t}=-\frac{\partial\delta B_z}{\partial x}-J_y,\;
\frac{\partial E_x}{\partial t}=-J_x .
\label{eq:maxwell_pic}
\end{equation}
The currents are deposited from the macroparticles using the cloud-in-cell
(CIC) scheme. The total longitudinal current contains the electron, ion,
and prescribed source contributions, $J_x=J_{xe}+J_{xi}+J_{x,b}$, whereas
$J_y=J_{ye}+J_{yi}$.

%
The moving source is represented by a prescribed Gaussian profile,
\begin{equation}
\rho_b(x,t)
=
\rho_{b0}
\exp\left[
-\frac{(x-x_b(t))^2}{2\sigma_b^2}
\right],
\qquad
x_b(t)=x_{b0}+V_b t,
\label{eq:pic_source}
\end{equation}
with source amplitude $\rho_{b0}$ and width $\sigma_b$. The associated
longitudinal source current is $J_{x,b}=V_b\rho_b$. Thus, the source acts as
an externally imposed moving driver rather than as a separately evolved
kinetic population.

The electromagnetic fields are defined on a staggered spatial grid and are
coupled to the particles through CIC interpolation and deposition. The
particle positions and velocities are advanced with the Boris algorithm,
after which the particle currents are deposited and the fields are updated.
%
The simulations employ $\Delta x=0.01$ and
$\Delta t=0.1\Delta x$, with 150 computational particles per cell for each
species. The initial particle velocities are sampled from Gaussian
distributions corresponding to the prescribed electron and ion thermal
velocities.

\subsection{Kinetic validation of the magnetosonic precursor}
\label{subsec:pic_mach_results}
To validate the nonlinear electromagnetic response obtained from the
two-fluid model, we perform corresponding PIC simulations for
the same subcritical and near-supercritical source velocities. The PIC
calculations use the same normalized plasma and source configuration
adopted in the fluid benchmark. The comparison is made for
$V_b=0.4c$, corresponding to $M=0.90$, and for
$V_b=0.505c$, corresponding to $M \simeq 1.14$, with
$V_{\rm ms}=0.4405c$. These cases directly test the transition from the
wake-dominated regime below the magnetosonic speed to the
precursor-dominated regime when the source velocity approaches and
exceeds the magnetosonic speed.

\begin{figure}[t]
    \centering
    \includegraphics[width=\columnwidth]{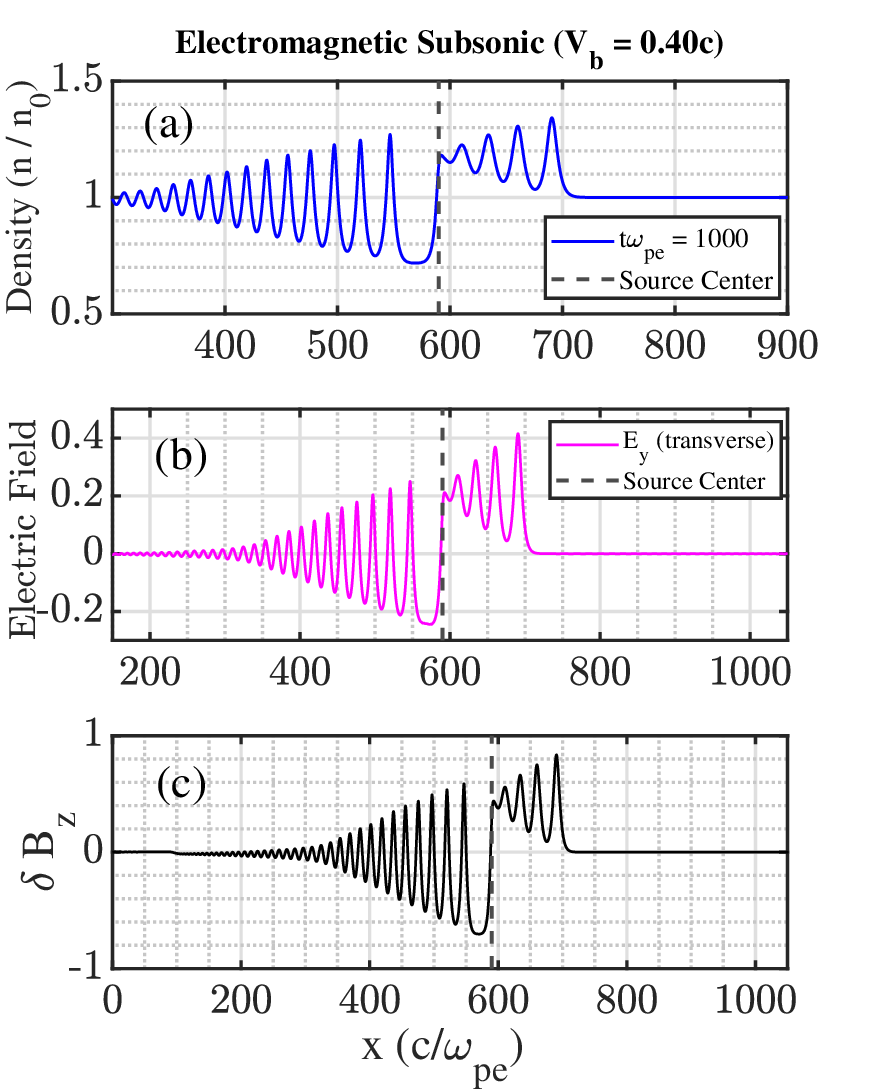}
    \caption{
    Kinetic PIC response for the subcritical source velocity
    $V_b=0.4c$ ($M=0.90$). The density, longitudinal electric field
    $E_x$, transverse electric field $E_y$, and magnetic-field
    perturbation $\delta B_z$ are shown. The vertical dashed line
    denotes the instantaneous source center. The predominantly
    wake-dominated response and weak forward-going structures are
    consistent with the subcritical two-fluid result shown in
    Fig.~\ref{fig:mach}(a,b).
    }
    \label{fig:PIC_subcritical}
\end{figure}

Figure~\ref{fig:PIC_subcritical} shows the PIC result for the
subcritical source velocity, $V_b=0.4c$. The density, transverse electric field $E_y$,
and magnetic-field perturbation $\delta B_z$ are shown as functions of
position. The vertical dashed line indicates the instantaneous source
center.
%
The PIC result exhibits a predominantly oscillatory wake response
behind the source. 
A relatively weak forward-going structure is also visible ahead of
the source. The corresponding structures are present simultaneously
in the longitudinal and transverse electric fields. 
This behavior is consistent with the
subcritical two-fluid result shown in Fig.~\ref{fig:mach}(a,b), where
the source velocity $V_b=0.4c$ is below the characteristic
magnetosonic speed, and the response is dominated by trailing wake
oscillations with only weak forward structures.
%
%
Thus, the kinetic calculation reproduces the principal electromagnetic
features predicted by the two-fluid model for the subcritical
regime. The additional density and magnetic-field diagnostics in the
PIC calculations show that these electric-field perturbations are
accompanied by corresponding plasma compression and magnetic
perturbations, providing a more complete kinetic description of the
electromagnetic wake.

\begin{figure}[t]
    \centering
    \includegraphics[width=\columnwidth]{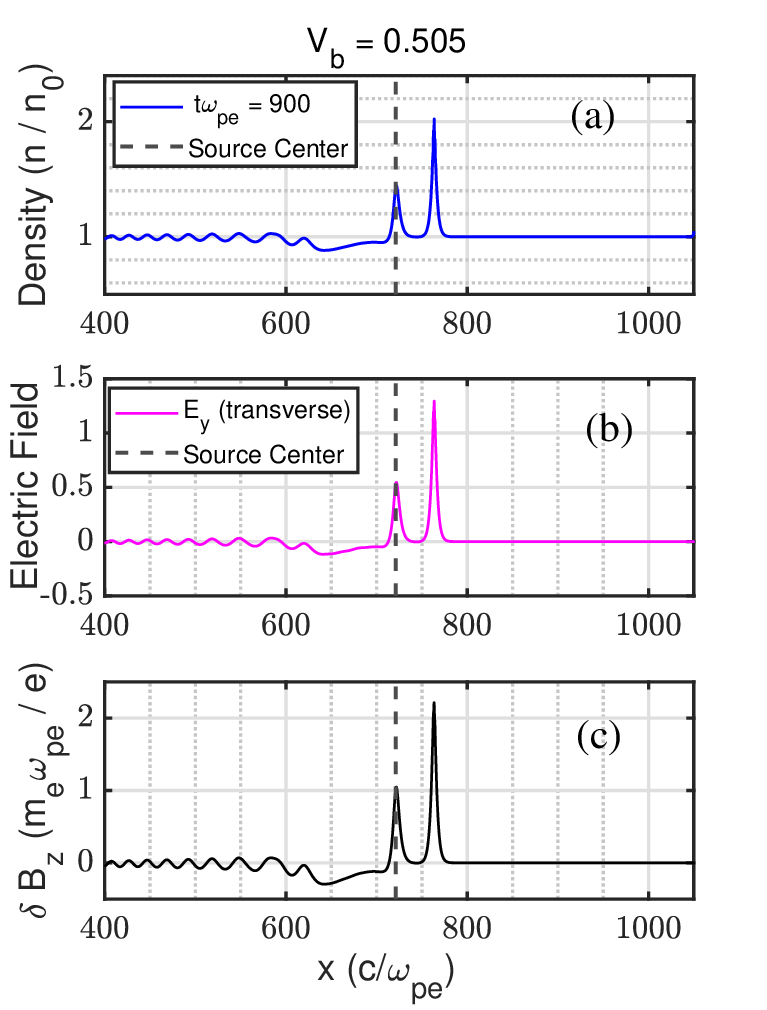}
    \caption{
    Kinetic PIC response for the near-supercritical source velocity
    $V_b=0.505c$ ($M\simeq1.14$). The density, longitudinal electric field
    $E_x$, transverse electric field $E_y$, and magnetic-field
    perturbation $\delta B_z$ are shown at $t\omega_{pe}=900$. The
    vertical dashed line denotes the instantaneous source center.
    Pronounced localized electromagnetic precursor structures develop
    near and ahead of the source, in agreement with the
    near-supercritical two-fluid result shown in
    Fig.~\ref{fig:mach}(c,d).
    }
    \label{fig:PIC_supercritical}
\end{figure}

The corresponding near-supercritical PIC result is shown in
Fig.~\ref{fig:PIC_supercritical} for $V_b=0.505c$ ($M\simeq 1.14$). This
case is compared with the near-supercritical two-fluid calculation
shown in Fig.~\ref{fig:mach}(c,d), for which $V_b=0.505c>V_{\rm ms}$.
With the same source velocities, both cases lie just
above the characteristic magnetosonic speed and therefore provide a
direct test of the transition identified by the fluid model.
A qualitative change in the plasma response is evident as the source
velocity crosses the magnetosonic threshold. In the PIC simulation,
the weak oscillatory response is replaced by pronounced localized
structures in the vicinity of and ahead of the source. The density
perturbation develops sharp compressive peaks with a maximum
normalized density of approximately $n/n_0\simeq2$. At the same
locations, strong localized structures appear in both $E_x$ (not presented here) and
$E_y$. The longitudinal and transverse electric fields, therefore,
remain spatially correlated, as predicted by the two-fluid model.
The magnetic-field perturbation $\delta B_z$ exhibits corresponding
localized peaks at the same positions, demonstrating that the
nonlinear structure contains a coupled magnetic response in addition
to the electric-field perturbations.
The comparison with Fig.~\ref{fig:mach}(c,d) shows that the principal
features of the near-supercritical fluid solution are reproduced by
the kinetic PIC calculation. In particular, the localized peaks in
both the longitudinal and transverse electric fields occur on the
forward side of the moving source and have comparable spatial
localization and characteristic amplitudes to those obtained from the
fluid calculation. The transition from the relatively weak
forward-going structures in the $V_b=0.4c$ case to the pronounced
localized structures for $V_b=0.505c$ is therefore consistently
reproduced by both models.


Overall, the comparison demonstrates that the source-velocity
dependence predicted by the two-fluid model is retained in the
kinetic description. For $M<1$, the response is predominantly
wake-dominated with only weak forward-going structures, whereas for
$M>1$ the response changes to pronounced localized structures ahead of
the source. The close correspondence of the $E_x$ and $E_y$
structures between the fluid and PIC calculations, together with the
additional density and magnetic signatures obtained from PIC,
provides kinetic validation of the electromagnetic magnetosonic
precursor predicted by the two-fluid model.

\subsection{Kinetic validation of the magnetosonic pinned }
\label{subsec:pinned}
The strongly supermagnetosonic regime provides a further test of the
two-fluid description through the formation of a source-locked
electromagnetic structure. We consider the same plasma and source
parameters used in the two-fluid calculation, with $B_0=2.5$ and
$V_b=0.70c$. Since the characteristic magnetosonic speed is
$V_{\rm ms}=0.4402c$, the corresponding Mach number is
$M\simeq1.59$.
Unlike the freely propagating precursor structures,
the nonlinear disturbance in this regime remains spatially localized
around the moving source. We therefore use the PIC calculation
to test whether this source-locked electromagnetic structure predicted
by the two-fluid model persists when the plasma response is described
in terms of particles.

\begin{figure}[t]
    \centering
    \includegraphics[width=\columnwidth]{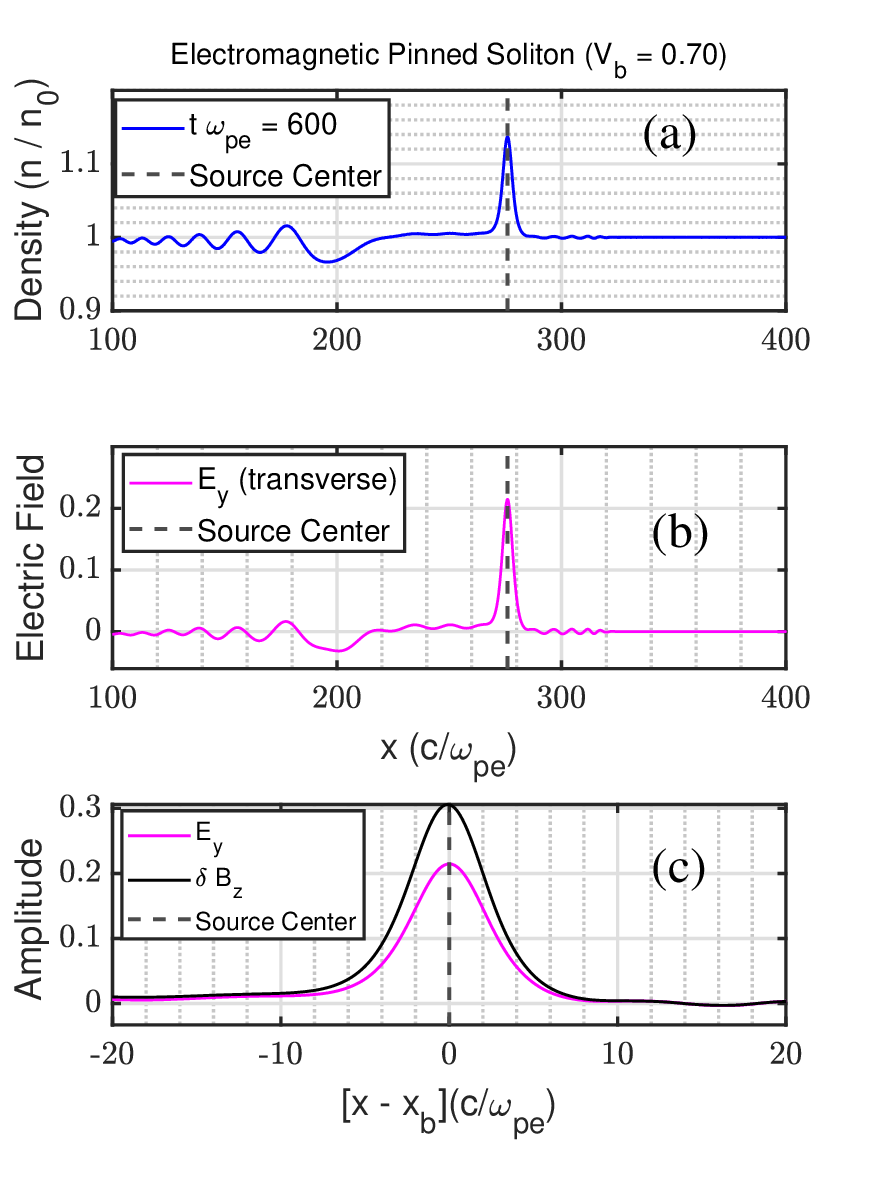}
    \caption{
    Kinetic PIC validation of the pinned electromagnetic structure for
    $V_b=0.70c$ ($M=1.59$). (a) Density perturbation at
    $t\omega_{pe}=600$; (b) transverse electric field $E_y$; and
    (c) source-frame profiles of $E_y$ and the magnetic-field
    perturbation $\delta B_z$ versus $x-x_b$. The localization of all
    perturbations around the source position demonstrates the
    source-locked character of the pinned structure.
    }
    \label{fig:pic_pinned}
\end{figure}

Figure~\ref{fig:pic_pinned} presents the corresponding PIC result.
Panel~(a) shows the density perturbation at
$t\omega_{pe}=600$. A pronounced localized density enhancement is
observed at the instantaneous source position, with a peak density of
approximately $n/n_0\simeq1.13$. The density maximum is centered very
close to the vertical dashed line marking the source position, while
only a comparatively weak oscillatory response extends away from the
source. Thus, the nonlinear compression remains localized to the source
rather than separating from it to form a freely propagating precursor.

The transverse electric field is shown in panel~(b). A narrow
positive peak in $E_y$ is located at essentially the same position as
the density enhancement and reaches an amplitude of approximately
$E_y\simeq0.21$ in the adopted normalization. The spatial coincidence
of the density and transverse electric-field maxima demonstrates that
the localized density compression is accompanied by a strong
electromagnetic response. The field rapidly decreases away from the
source, further emphasizing the localized character of the structure.

A direct comparison in the source frame is given in panel~(c), where
the horizontal coordinate is $x-x_b$. Both $E_y$ and the magnetic-field
perturbation $\delta B_z$ are centered near $x-x_b=0$, i.e., at the
source location. The magnetic perturbation has a broader spatial
profile and a larger peak amplitude, reaching approximately
$\delta B_z\simeq0.30$, while $E_y$ reaches approximately
$0.21$. The two quantities therefore form a coherent, spatially
localized electromagnetic structure that remains centered on the
moving source.

The kinetic result is in close agreement with the corresponding
two-fluid solution shown in Fig.~\ref{fig:fluid_pinned}.
%
%
The PIC calculation reproduces the characteristic features of these structures to within the
resolution of the present comparison. More importantly, the
source-frame profiles of $E_y$ and $\delta B_z$ in
Fig.~\ref{fig:pic_pinned}(c) closely follow the localized shapes
obtained from the two-fluid calculation. Thus, the agreement is not
restricted to the existence of a density peak; the coupled
electromagnetic structure itself is recovered in the kinetic model.
%
The recovery of
the approximately same source-localized density and electromagnetic structures in the
PIC calculation, therefore, provides a kinetic validation of the pinned
state predicted by the fluid model.
\section{Conclusions}
\label{sec:conclusion}
We have developed a one-dimensional two-fluid framework for studying
nonlinear plasma structures generated by a localized moving charged source
and have used an independent kinetic particle-in-cell model to test the
principal fluid predictions. The formulation retains the electron and ion
fluid dynamics together with the longitudinal and transverse
electromagnetic fields, thereby providing a common description of the
electrostatic and magnetized responses. The prescribed Gaussian source also
allows the nonlinear response to be examined in a controlled manner while
maintaining a direct connection with earlier charged-source models.

The zero-magnetic-field calculation recovers the expected electrostatic
ion-acoustic precursor response. The localized density structures satisfy,
to a good approximation, the characteristic KdV amplitude--width relation
\(A L^2\simeq\mathrm{const.}\), providing a useful diagnostic of their
solitary-wave character. This benchmark establishes the numerical
consistency of the two-fluid formulation with the established precursor
picture and provides the baseline against which the finite-magnetic-field
results can be interpreted.

In the magnetized regime, two complementary diagnostics identify the linear
character of the far wake. The dominant spectral component occurs at
\(\omega_{\rm peak}\simeq0.0776\,\omega_{pe}\), while direct tracking of a
far-wake density maximum yields a propagation velocity
\(v_{\rm ph}\simeq0.4255c\). The measured velocity is close to the
characteristic fast-magnetosonic speed \(V_{\rm ms}\simeq0.4402c\) for the
adopted normalization. The far-wake therefore provides a linear
magnetosonic reference against which the subsequent nonlinear structures
can be compared.

The nonlinear simulations show a clear change in the response with source
velocity. For \(V_b<V_{\rm ms}\), the fields are dominated by trailing
oscillatory wake structures, with only weak forward perturbations. When the
source velocity slightly exceeds the magnetosonic speed, localized
structures develop ahead of the source and appear simultaneously in the
longitudinal and transverse electric fields. Their electromagnetic
character is further supported by the approximately invariant
amplitude--width product of the leading structure. At higher
magnetosonic velocity, the dominant nonlinear perturbation becomes
locked to the source trajectory. The coincidence of the density maximum,
transverse electric-field peak, and co-moving space--time ridge identifies
this state as a pinned electromagnetic structure.

The kinetic PIC calculations reproduce the principal features of the
two-fluid solution. In the subcritical regime, the PIC model recovers the
wake-dominated response and weak forward-going structures. In the
near-supercritical regime, it reproduces the emergence of localized
density, longitudinal electric-field, transverse electric-field, and
magnetic-field perturbations near and ahead of the moving source. The
kinetic calculation also recovers the source-localized electromagnetic
structure in the higher-velocity pinned regime. The agreement between the
two descriptions is significant because the underlying representations
are fundamentally different: the fluid model evolves macroscopic moments,
whereas the PIC model obtains the response from the self-consistent motion
and current deposition of individual computational particles. The
persistence of the same spatially correlated nonlinear signatures across
these descriptions therefore provides a useful kinetic validation of the
fluid predictions.

The developed and kinetically validated two-fluid--Maxwell model provides a foundation for investigating source-driven plasma dynamics beyond the regimes considered here. A natural next step is to derive a forced Korteweg--de Vries (fKdV) equation from the present framework under appropriate weakly nonlinear and weakly dispersive conditions. By retaining the leading effects of nonlinearity, dispersion, and source forcing in a reduced description, the fKdV model offers a computationally efficient approach to systematic parameter studies of magnetosonic precursor and pinned structures, with its predictions assessed against the full two-fluid results. This reduced model also provides a basis for exploring realistic ionospheric parameters, including plasma density, temperature, ion composition, magnetic-field strength, and source velocity, to determine how these factors influence the formation and evolution of nonlinear structures. Extending the full two-fluid--Maxwell framework beyond its present one-dimensional geometry would also allow examination of the influence of oblique source trajectories and magnetic-field orientations. Moreover, the separate evolution of electron and ion fluids provides a natural basis for incorporating time-dependent source charging through electron and ion collection currents. Coupling such a charging model to the plasma solver, potentially using a particle-based representation of the charged object, would enable investigation of how source-charge fluctuations modify the generation and propagation of plasma waves. Finally, the retention of species-resolved dynamics and electromagnetic-field evolution makes the framework a promising starting point for studying additional wave modes, including Alfv\'en and whistler, provided that the geometry, field components, and numerical resolution are extended to resolve their relevant dispersion and timescales. Together, these developments can broaden the application of the present framework from the validated magnetosonic response to a wider range of source-driven phenomena in space plasmas.
\section*{Conflicts of Interest}
The authors declare no conflicts of interest.
\section*{Data Availability Statement}
The data that support the findings of this study are available from the
corresponding author upon reasonable request.
\section*{Acknowledgments}
This work is supported by the Department of Space, Government of India. 
\appendix
\section{Linear Dispersion Benchmark}
\label{app:linear_dispersion}

\begin{figure}[t]
    \centering
    \includegraphics[width=1.0\columnwidth,height=6cm]
    {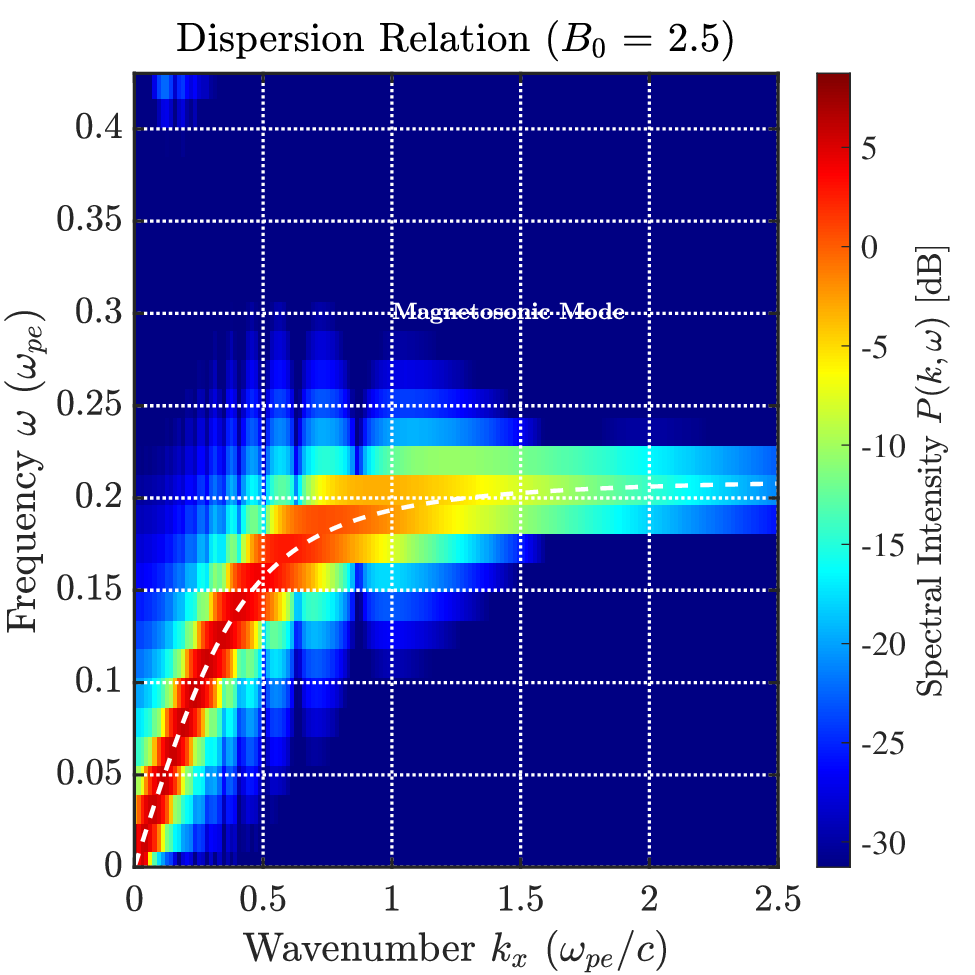}
    \caption{Low-frequency dispersion benchmark for
    $m_i/m_e=25$ and $B_0=2.5$. The color map shows the
    space--time Fourier spectrum of $\delta B_z$, and the dashed
    curve is the analytical magnetosonic branch obtained from
    Eq.~\eqref{eq:app_dispersion}.}
    \label{fig:linear_dispersion}
\end{figure}

The analytical dispersion relation is derived by linearizing the
two-fluid equations coupled to Maxwell's equations, presented in
Sec.~\ref{sec:fluidmodel}, about a homogeneous equilibrium and
applying the standard cold-plasma dielectric-tensor formulation
\cite{Stix1992}. For perpendicular propagation,
$\mathbf{k}=k_x\hat{\mathbf{x}}\perp\mathbf{B}_0$, the resulting
linear system is
\begin{equation}
\begin{pmatrix}
S & iD\\
-iD & S-N^2
\end{pmatrix}
\begin{pmatrix}
\delta E_x\\
\delta E_y
\end{pmatrix}=0,
\qquad
N^2=\frac{k_x^2c^2}{\omega^2},
\end{equation}
where
\begin{align}
S(\omega)&=1-\frac{\omega_{pe}^2}{\omega^2-\Omega_{ce}^2}
-\frac{\omega_{pi}^2}{\omega^2-\Omega_{ci}^2},\\
D(\omega)&=\frac{\Omega_{ce}}{\omega}
\frac{\omega_{pe}^2}{\omega^2-\Omega_{ce}^2}
-\frac{\Omega_{ci}}{\omega}
\frac{\omega_{pi}^2}{\omega^2-\Omega_{ci}^2}.
\end{align}
Here, $\Omega_{ce}$ and $\Omega_{ci}$ are positive cyclotron-frequency
magnitudes. Requiring a non-trivial solution gives the dispersion
relation
\begin{equation}
\frac{k_x^2c^2}{\omega^2}
=\frac{S^2-D^2}{S}.
\label{eq:app_dispersion}
\end{equation}
The benchmark uses the same normalized mass ratio and background
magnetic field as the main simulations, $m_i/m_e=25$ and $B_0=2.5$,
with the pressure terms set to zero for the cold-limit calculation.
Thus, $\omega_{pe}=1$, $\omega_{pi}=0.2$,
$\Omega_{ce}=2.5$, and $\Omega_{ci}=0.1$. The low-wavenumber
magnetosonic phase speed follows from Eq.~\eqref{eq:app_dispersion}:
$V_{\mathrm{ms}}
=\lim_{k_x\to0}\frac{\omega}{k_x}
=\frac{c}{\sqrt{S(0)}}\simeq0.4402c.$
%
To construct Fig.~\ref{fig:linear_dispersion}, a small-amplitude
broadband perturbation is applied to the source-free numerical
equation presented in
Sec.~\ref{sec:fluidmodel} and the magnetic-field perturbation
$\delta B_z(x,t)=B_z(x,t)-B_0$ is recorded. 
The spectral intensity $P(k_x,\omega)$ is plotted,
with the analytical low-frequency magnetosonic branch
superimposed using the cold-plasma dispersion relation
$k_x(\omega)=\frac{\omega}{c}
\sqrt{\frac{S^2-D^2}{S}}$,
restricted to the propagating region where
$\frac{S^2-D^2}{S}>0.$
As shown in Fig.~\ref{fig:linear_dispersion}, the analytical curve
follows the principal low-frequency spectral ridge over the resolved
wavenumber range. Although the full dispersion relation also admits
higher-frequency electromagnetic branches, only the low-frequency
range and the magnetosonic branch are emphasized here, since they
are directly relevant to the wake dynamics investigated in this work.
The wake diagnostics in Sec.~\ref{subsec:wake} provide a complementary
comparison with the source-driven simulation. The measured spectral
peak, $\omega_{\mathrm{peak}}\simeq0.0776\,\omega_{pe}$, and the
tracked phase velocity, $v_{\mathrm{ph}}\simeq0.4255c$, yield
\begin{equation}
k_{\mathrm{eff}}=\frac{\omega_{\mathrm{peak}}}{v_{\mathrm{ph}}}
\simeq0.1824\,\frac{\omega_{pe}}{c},
\qquad
\lambda_{\mathrm{eff}}=\frac{2\pi}{k_{\mathrm{eff}}}
\simeq34.45\,\frac{c}{\omega_{pe}}.
\end{equation}
The measured phase velocity is close to the cold-limit value
$V_{\mathrm{ms}}\simeq0.4402c$. At the inferred wavenumber, the
analytical relation predicts $\omega\simeq0.0760\,\omega_{pe}$,
approximately $2.1\%$ below the measured spectral peak. This
agreement supports the magnetosonic interpretation of the wake.
The small difference is consistent with comparing a driven wake,
which may retain finite-amplitude effects, with a linear,
homogeneous, source-free cold-plasma benchmark.
\bibliographystyle{apsrev4-2}
\bibliography{paper1}
\end{document}